\documentclass[aps,prl,reprint,groupedaddress,twocolumn]{revtex4}
\usepackage{graphicx}   
\usepackage{amsmath,amssymb,wasysym}
\usepackage[amsmath,thmmarks,framed]{ntheorem}

\usepackage{color}

\usepackage{ulem}

\begin{document}


\title{ Magnetic induction measurements using an all-optical $^{87}$Rb atomic magnetometer}

\author{Arne Wickenbrock }
\affiliation{Department of Physics and Astronomy, Gower Street, London WC1E 6BT, United Kingdom}
\email[]{A.Wickenbrock@ucl.ac.uk}
\author{Fran\c{c}ois Tricot}
\affiliation{Department of Physics and Astronomy, Gower Street, London WC1E 6BT, United Kingdom}
\author{Ferruccio Renzoni}
\affiliation{Department of Physics and Astronomy, Gower Street, London WC1E 6BT, United Kingdom}

\date{\today}

\begin{abstract}
In this work we propose, and experimentally demonstrate, the use of a self-oscillating all-optical atomic
magnetometer for magnetic induction measurements. Given the potential for miniaturization 
of atomic magnetometers, and their extreme sensitivity, the present work shows that atomic magnetometers may play a key role in the development of instrumentation for
magnetic induction tomography.

\end{abstract}


\maketitle



Magnetic Induction Tomography (MIT) \cite{mit} allows the detetection and characterization of 
conductive  objects. In an MIT set-up an ac magnetic field induces eddy currents in the conductive 
object of interest. The eddy currents generate an additional magnetic field, whose detection 
gives access to the magnetic and electric properties of the object. Spatial resolved 
measurements then allow the reconstruction of the image of the object. MIT systems 
typically use coils of wire for the detection. The finite dimension of the sensor limits the 
spatial resolution of the system, which is a major drawback of the technique.

In this work we propose, and experimentally demonstrate, the use of all-optical atomic magnetometers for magnetic induction measurements. Atomic magnetometers
\cite{budker_review} are increasingly popular due to their extreme sensitivity, which has  recently been shown to surpass the sensitivity of SQUIDs. They can measure
extremely small static \cite{romalis2003,romalis2013}, as well as oscillating \cite{romalis2005,gawlik2012}, magnetic fields. Here we demonstrate 
that they can also measure the phase variation in an ac magnetic field induced by the establishment of eddy currents in a conductive object placed in the proximity of the sensor. Given the potential for miniaturization 
of atomic magnetometers, and their extreme sensitivity, the present work shows that atomic magnetometers may play a key role in the development of instrumentation for
magnetic induction tomography.

In all-optical magnetometers  an atomic sample plays the role of the sensor.
The atomic vapor is spin polarized with a strong on-resonant pump beam. When the pump beam is switched off the collective magnetic moment of the atoms start to precess with the Larmor frequency around the magnetic field vector. The precession can be probed by a weak probe beam, with linear polarization, via a
process called nonlinear magneto-optic rotation (NMOR) \cite{budker2002}. The  resulting oscillating polarization  of the  probe beam allows one to directly measure the Larmor frequency, and thus to derive the magnetic field amplitude. An important  issue in atomic magnetometers is the decay of the atomic spin, which is a limiting factor for the sensitivity.  Modulating the pump beam periodically with the Larmor frequency either in frequency or in amplitude  allows to compensate the spin decay by realigning the atomic magnetic moment along the beam axis. In a self-oscillating setup \cite{schwindt,belfi}, the polarization rotation of the probe beam itself is used to modulate the pump beam in a feedback circuit causing the system to oscillate, in the limit of small spin decoherence, at the Larmor frequency, which can then be directly read out using a frequency meter or a spectrum analyzer.\\

In this work we use the high-bandwidth of an all-optical $^{87}$Rb  magnetometer to detect an applied oscillating magnetic field. Sinusoidal oscillations of up to 50\,kHz can be detected with an amplitude of around 100\,$\mu G$. 
An oscillating magnetic field induces eddy currents in conducting objects, which itself cause a magnetic field oscillating with the same frequency. The combined field is measured at the position of the sensor, in our case the $^{87}$Rb vapor. Using a dual phase lock-in amplifier the phase of the measured signal can be compared to the phase of the driving. This contains information about the induced current and therefore about the presence and nature of conducting objects close to the sensor. We detect different objects placed close to the cell by observing the phase for different frequencies of the driving using an all-optical magnetometer.

\begin{figure*}
	\centering
	\includegraphics{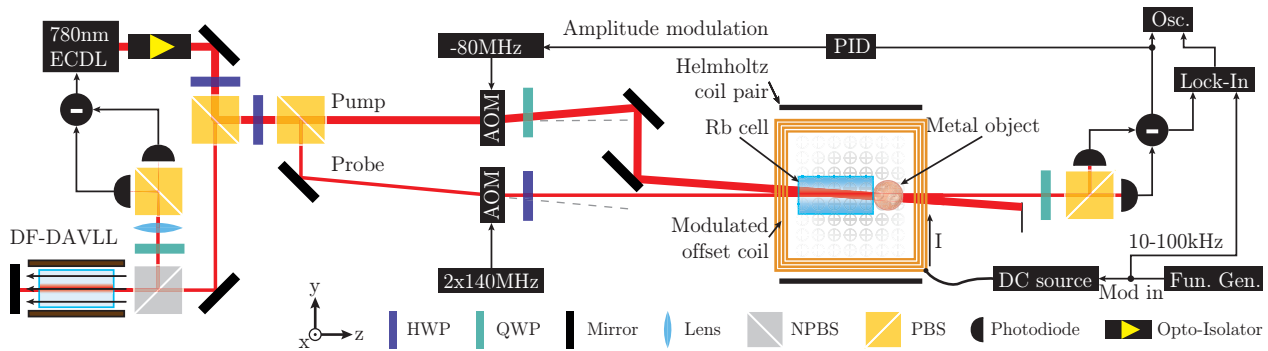}
	\caption{Schematic of the atomic magnetometer for magnetic induction measurements: 
           the laser beam produced by an extended cavity diode laser lasing at 780\,nm is split into three beams. 
           One part is used for a Doppler free dichroic atomic vapour laser lock. The other two beams are the pump and 
           the probe beam for the self-oscillating magnetometer. The probe is shifted up in frequency with a double pass 
           AOM by 280\,MHz and linearly polarized while the circular polarized pump is shifted down by another AOM 
          (-80\,MHz). Both beam intersect at a small angle of around 1.5 degrees in the vapor cell. The polarization 
           rotation of the probe is measured using a balanced polarimeter after the cell. In self-oscillation mode this 
          signal is then fed back to drive the intensity of the pump AOM. The rubidium cell is positioned in the 
          center of a rectangular coil capable of producing a homogenous magnetic field perpendicular to the probe 
          beam propagation direction. This coil is also used to produce an additional oscillating magnetic field 
         component, which induces eddy currents in conducting materials close to the sensor. Those currents
          result in a phase change of the measured oscillating magnetic field, which can be detected with a lock-in 
          amplifier. When the magnetic induction is measured the objects are placed close to the interaction region 
          on a plastic pedestal.}
	\label{fig:Figure1}
\end{figure*}
The experimental apparatus is shown schematically in figure \ref{fig:Figure1}. 
The magnetic sensor is a 5\,cm vapor cell filled with the naturally occuring mixture of $^{85}$Rb and $^{87}$Rb. 
The cell is coated with  polydimethylsiloxane (PDMS) and filled with 5 Torr of Argon gas. The cell is heated to $70\,^{\circ}\mathrm{C}$ 
to increase the amount of rubidium in the vapor phase. The heat is generated by a current driven resistor 30\,cm 
away from the cell and transported via heat pipes to the vapor. The cell is placed in a rectangular solenoid providing an 
offset magnetic field in the $z$ direction. The current through the coil can be modulated, with a modulation frequency up to 50\,kHz. In this
way, it is possible to introduce an ac magnetic field component to the offset field. Another set of Helmholtz coils along the 
$y$ direction is used to compensate the magnetic field in this direction. Apart from 
these coils, no additional effort is made to shield or compensate magnetic field gradients or noise. This obviously 
limits the atomic coherence lifetime, and thus the sensitivity of the magnetometer. However, this is not relevant to the
present work, which aims to demonstrate the possibility to perform magnetic induction measurements.
Close to the cell is a pedestal to place conducting objects for the magnetic induction measurement.
The laser beam is generated by a a home built extended cavity diode laser (ECDL) lasing at 780\,nm. 
This corresponds to the D$_2$ line of both rubidium isotopes.
A fraction of the light is used for a Doppler-free dichroic atomic vapor laser lock (DF-DAVLL) setup to stabilize the 
laser frequency to the $^{87}$Rb $D_2$ $F=2 \rightarrow F'=2,3$ crossover transition.
The remaining laser beam is then split in two on a polarizing beam splitter, to form the pump and probe beam.
 The probe beam is then shifted up in frequency by 280\,MHz using a acousto-optical modulator (AOM) in double pass 
configuration while the pump beam is shifted down in frequency by 80\,MHz, resulting in an overall blue detuning of 360\,MHz 
of the probe beam with respect to the pump. The AOM in the pump beam path is also used to control the pump's intensity and 
enables therefore the operation of the magnetometer in a self-oscillating mode. The polarization of the two beams is then
prepared by quarter- and half-waveplates. Before entering the cell, the pump beam is circularly polarized, with a beam waist 
of $\phi=2.1$\,mm and a power P = 0.95\,mW,  corresponding to a peak intensity of 15.2\,mW/cm$^2$ or 9.2\,$I_{\text{Sat}}$. The probe beam is linearly polarized, with a diameter $\phi=2.1$\,mm and a power 
P = 1.1\,mW.  The beam intensities were optimized within the experimental constraints for maximum signal to noise, but the self-oscillating magnetometer is fairly insensitive to variations of the pump power, as long as it is above saturation intensity.

The two beams are overlapped, forming a small angle of 1.5\,degrees. 
After propagation through the cell, the polarization rotation of the linear polarized probe beam is measured with 
a balanced polarimeter,  while the pump beam is blocked. In self-oscillating mode, the signal of the polarimeter is used to drive the intensity of the pump AOM via a PID controller (SRS SIM900). The PID controller is necessary to achieve self-oscillations which appear for a wide range of gain settings. The controller is also used to add an offset to the output to ensure that the intensity response of the AOM driver is linear. In closed loop configuration the system oscillates at the frequency with the strongest gain, so that in general the oscillation frequency depends weakly on the gain settings (and the delay times and frequency characteristics of the other components in the loop.)

\begin{figure}
	\centering
		\includegraphics[width=0.45\textwidth]{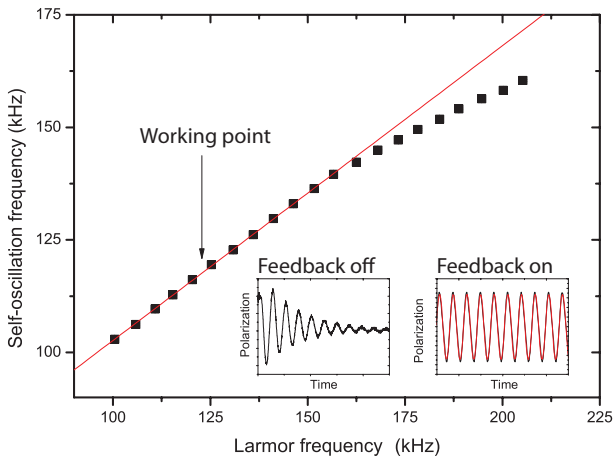}
	\caption{Self-oscillation frequency as a function of the Larmor frequency for different magnetic offset fields. Around the working point the self-oscillation frequency depends linear on the Larmor frequency. The insets show the response of the probe polarization to a sudden switch off of the pump (left) and to the polarization signal being used to drive the pump's intensity via the AOM via a feedback circuit (SRS SIM960)}.
	\label{fig:Figure2}
\end{figure}

We performed a first series of measurements to calibrate the self-oscillating magnetometer. For an applied DC magnetic field, and the magnetometer in open loop configuration (no feedback), the Larmor frequency was measured by observing the decaying polarization oscillation occuring after the pump is switched off. For different offset magnetic fields, the experimental data were fitted by a sine function with an exponential decaying amplitude, so to extract the Larmor frequency. For each value of the dc offset field, the system was then switched to the self-oscillating mode and the frequency recorded with a spectrum analyzer. Figure \ref{fig:Figure2} displays the frequency of oscillation of the magnetometer as a function of the Larmor frequency corresponding to the applied magnetic field. Around 125\,kHz the self-oscillation frequency exhibits a linear relationship to the Larmor frequency and can be used to measure the magnetic field. Thus this frequency was chosen as the working frequency of our magnetometer, so that the self-oscillation occured in the middle of the linear region. The insets in figure \ref{fig:Figure2} show an example of the polarization decay after the pump beam is switched off (left) and the self-oscillating response for the same magnetic field (right) when the polarization is fed back to the RF intensity of the pump.

\begin{figure}
	\centering
		\includegraphics[width=0.45\textwidth]{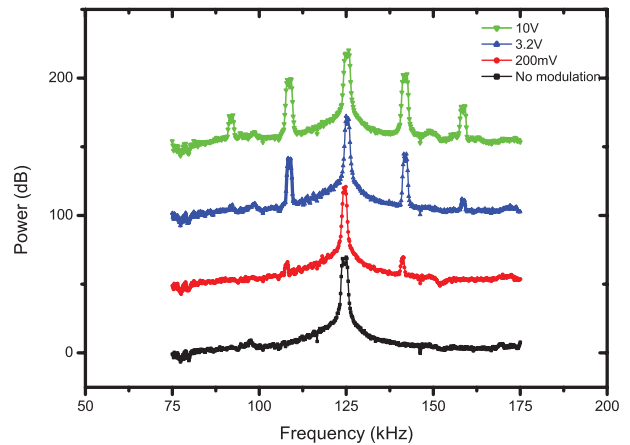}
	\caption{Power spectrum of the polarization oscillation signal in self-oscillating mode for different amplitudes of an applied 10\,kHz modulation of the offset magnetic field. This corresponds to a frequency modulation of the Larmor frequency which produces sidebands in the spectrum. The traces are offsetted by 80\,dB to each other for visibility. The modulation depth decreases from top to bottom.}
	\label{fig:Figure3}
\end{figure}
We note here the fast decay of the polarization as it can be observed in the left inset of figure \ref{fig:Figure2}. Spin relaxation times of several ms are achievable in vapour cells with buffer gas, while in the presented experiment the collective spin 1/e-decay time is below $20\,\mu$s. We attribute this to the lack of magnetic field gradient compensation and the presence of time varying magnetic fields, since the experiment is conducted in a completely unshielded laboratory environment.

\begin{figure}
	\centering
		\includegraphics[width=0.45\textwidth]{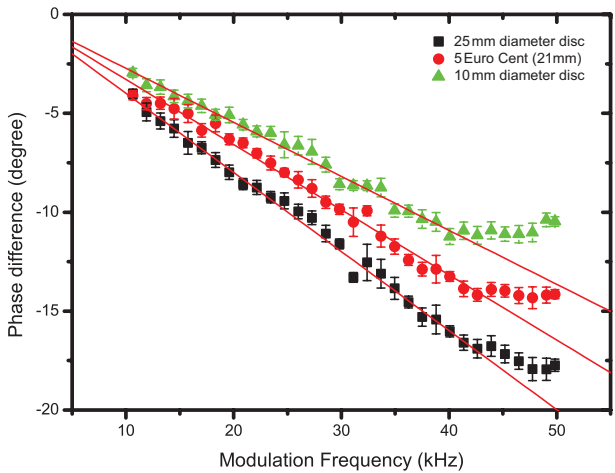}
	\caption{Induction phase measurement for different modulation frequencies and different objects.
	The phase of the modulation signal as read out via the lock-in amplifier. The phase change indicates the presence of a conducting object close (1.5\,cm) to the interaction region of pump and probe and exhibits a linear relationship to the frequency of the modulation. The measurement was repeated for three different items, a 25\,mm copper disc, a 5\,Euro Cent coin and 10\,mm copper disc, which can be distinguished in such a measurement.}
	\label{fig:Figure4}
\end{figure}

In a next step we applied a small sinusoidal amplitude modulation of the offset coil current, which corresponds to frequency modulation of the Larmor frequency. A power spectrum of the self-oscillating signal for different amplitudes of an applied 10\,kHz modulation can be seen in figure \ref{fig:Figure3}. An investigation of the central peak confirms the presence of spurious time varying fields. Especially several power line related 50\,Hz sidebands are resolvable. In fact the low frequency ac modulation of the magnetic field background is so strong that the spectrum effectively splits into two components seperated by around 1\,kHz. This explains the width of the central feature as seen in figure \ref{fig:Figure3} but does not affect the magnetic induction measurement (due to the comparable small frequency with respect to the applied modulation).
For increasing amplitude, sidebands at multiples of the modulation frequency appear. The smallest applied modulation amplitude where  sidebands could be detected corresponded to a magnetic field variation of just 100$\mu$G (at 10\,kHz). 
The measured frequency oscillation was then compared to the initial driving signal using a dual-phase lock-in amplifier with the capacity to display phase and magnitude of the signal. Placing an object in close proximity to the cell resulted in a variation of the phase of the measured signal. To study the behaviour of the phase as a function of the driving frequency for different objects the phase was firstly recorded with no object present. The phase signal was directly output onto a digital oscilloscope and the modulation frequency swept from 10\,kHz to 50\,kHz within 100s. Four traces were averaged to improve the signal to noise ratio. Then different conducting objects were placed close to the cell and the procedure repeated. At last the phase without any objects was subtracted from the phase traces including the objects. The results after offset subtraction are shown in figure \ref{fig:Figure4}.

The behaviour of the phase change appears linear with the modulation frequency and depends on the size of the object. This is in agreement with measurements of magnetic induction performed with conventional sensors made of a coil of wire. The largest phase change occured for a copper coin with 25\,mm diameter and a thickness of 1\,mm oriented perpendicular to the offset magnetic field at a distance of 1.5\,cm away from the interaction region along the direction of the probe and slightly above the beam  (4\,mm from the center of the probe, so that it can pass unperturbed underneath the coin).

In conclusion we demonstrated the use of an all-optical self-oscillating magnetometer with amplitude modulation for magnetic induction measurements.
We detected the phase of an applied oscillating magnetic field, and demonstrated the possibility to detect phase variations as a result of eddy currents 
established in a conductive object placed in the proximity of the sensor. Given the potential for miniaturization of atomic magnetometers, and their extreme sensitivity, the present work 
may open a new avenue in the development of instrumentation for magnetic induction tomography. The present work presented measurements in the 100 kHz range, thus
of direct relevance ot the detection, characterization and imaging of metallic object. The technique can also be extended to the MHz range which is of direct 
relevance to the imaging of biological objects \cite{bio}.

\begin{acknowledgments}
 This work was supported by a Marie Curie International Research Staff Exchange Scheme Fellowship 
within the 7th European Community Framework Programme.
We would like to thank Yordanka Dancheva for useful discussions, and Emilio Mariotti for providing the 
Rubidium cell used in this work.
\end{acknowledgments}.


\end{document}